# Spectral Features of the Fourth Order Irreducible Correlations in a Monolayer Semiconductor

Jiacheng Tang,[1,2] Cun-Zheng Ning[2,1]*

[1]Department of Electronic Engineering, Tsinghua University, Beijing 100084, China
[2]College of Integrated Circuits and Optoelectronic Chips, Shenzhen Technology University, Shenzhen 518118, China

* Corresponding author. Email: ningcunzheng@sztu.edu.cn

**Abstract**

**Understanding high-order correlations or multi-particle entities in a many-body system is not only of fundamental importance in condensed matter physics, but also critical for many technological applications. So far, higher-order multi-particle irreducible correlations in semiconductors have not been studied beyond the second-order or two-particle case. In this paper, we study the correlation of two electrons and two holes (2e2h) using the four-body Bethe-Salpeter equation (4B-BSE) and applied to the calculation of the helicity-resolved absorption between the two-body and four-body states for a monolayer $MoTe_2$. Surprisingly, we found a rich series of spectral peaks within an energy span of ~40 meV below the exciton that has not been seen before. To understand the origin of the new spectral peaks, the Feynman diagrams of the 4B BSE are recast into the cluster expansion formalism, allowing us to study the individual effects of selected clusters or correlations of various orders. We found that the irreducible clusters of orders up to the 3$^{rd}$ and their factorized combinations cannot explain the spectral features. Importantly, we found that the 4$^{th}$ order irreducible correlation is necessary and sufficient to explain the new features. The 4$^{th}$ order irreducible correlation corresponds to a four-particle irreducible cluster involving two electrons and two holes, alternatively called quadron or quadruplon. The new 4$^{th}$ order correlation or four-particle entity not only enriches our understanding of many-body correlations but also could provide new mechanism for light emission or absorption for possible new optoelectronic devices.**

Many-body complexes such as one-, two-, and three-particle entities such as electrons (*e*)/holes (*h*), excitons (X), trions (T), and their product states are known and widely explored, higher order irreducible correlated entities such as four-particle states are much less understood, especially beyond bi-excitons (BX) in semiconductors. Among few existing examples[1-7] in physics that have been studied either theoretically or experimentally, the most closely-related analogy is tetraquarks[2-5] vs. meson-molecules[5,8] in elementary particle physics. Few-body versions of the Bethe-Salpeter Equations (BSEs) or Semiconductor Bloch Equations (SBEs) were formulated beyond the well-known two-body case[9-12] to calculate the few-particle composites such as three- or four-quark species[8,13], two-electron-one-hole ($2e1h$, vice versa) three-body (3B) trion states[14-21], and exciton polarons[22-24] such as the Suris tetron[25] *etc*. For the $2e2h$ four-body (4B) system, only the special case of bi-exitons[26-29] (BX) or bi-excitons fine structures[30,31] (BXFS) have been studied so far. Other related examples are the 4-body quadrupling[32,33] reported for superconductors, without addressing the issue of the irreducibility.

Given the existence of many unique features such as being direct bandgap semiconductors[34,35], large excitonic binding energies[34,35] (for also trions[36-43], bi-excitons[30,44-48], and charged bi-excitons[49-54]), and spin-valley locking[55,56], monolayer transition metal dichalcogenides (ML-TMDCs) provide an important platform for the study the multi-particle excitonic complexes or higher order correlations.

In this letter, we conducted a theoretical investigation into the possible existence of the 4-body irreducible correlations and the general spectroscopic features of the four-body states, beyond those known lower-order species and their production states. Solving the 4B-BSE for the $2e2h$ system and the 2B-BSE for an $eh$ system, we are able to calculate the dielectric function related to the optical transitions between the 2-body and 4-body states, which correspond to the excited state absorption (ESA). In contrast to the relatively simple spectrum seen in typical experimental measurement, our calculation shows a rich variety of spectral peaks in a wide range of photon energies for ML-$MoTe_2$. More specifically, up to six spectral peaks were obtained, extending over 40 meV from below T all the way up to X. To understand the new spectral features more intuitively, we re-arrange the Feynman diagrams of the 4B-BSE such that their correspondence with the cluster expansion can be established. The correspondence allows us to identify the spectral contributions of the irreducible clusters of various orders up to the 4$^{th}$. Interestingly, we show that the clusters up to the third order, corresponding to the trions and bi-excitons cannot produce most of the new spectral features, thus excluding the trions and bi-excitons as their origin. Importantly, our results show that the 4$^{th}$-order irreducible 2e2h-cluster, which correspond to the quadron or quadruplon, is necessary and sufficient in producing all the sequence of spectral features, thus providing spectroscopic signatures of the existence of the fourth order irreducible correlation or the quadruplon.

We describe the single-particle band structure of ML-MoTe$_2$ near the K and K' points by using the gapped two-band massive Dirac model[55]. Inserting parameters obtained from the DFT calculations into the model, one obtains analytic expressions for the electronic bands, the eigen wavefunctions, and the inter-band dipole matrix elements[57]. More details can be found in Supplementary Information (SI) S1.

The wavefunctions of the 4B and 2B states, $|e_1h_1e_2h_2\rangle$ and $|e_3h_3\rangle$, can be expressed in terms of creation and annihilation,

$$|e_1h_1e_2h_2\rangle = \sum_{(v_1,c_1,v_2,c_2)} B^{v_1,c_1,v_2,c_2} \hat{a}^\dagger_{c_1} \hat{a}^\dagger_{c_2} \hat{a}_{v_1} \hat{a}_{v_2} |0\rangle \qquad (1)$$

$$|e_3h_3\rangle = \sum_{(v_3,c_3)} A^{v_3,c_3} \hat{a}^\dagger_{c_3} \hat{a}_{v_3} |0\rangle \qquad (2)$$

where $v_i$'s and $c_i$'s index the valence (v) and conduction (c) band single-particle states, respectively, including the respective momenta that are discretized using a Monkhorst-Pack mesh. Coefficients $A^{v_3,c_3}$ and $B^{v_1,c_1,v_2,c_2}$ are determined by the 2B-BSE and 4B-BSE, respectively, as shown in eq. (1) & (2) in SI S2.

FIG. 1(a) shows the diagrammatic representation of such a 2e2h 4B-BSE (eq. (2) in SI S2). As can be seen, the structure of 4B-BSE is an integral equation similar to that of 2B-BSE or a one-particle Dyson's equation[58,59]. The dashed-box in FIG. 1(a) represents the complete e-e, h-h, and e-h interaction kernels. For the sake of brevity, we detailed in FIG. 1(b) only the screened interaction kernels (see also Ref. [60,61]). Diagonalizing the 4B-BSE and the 2B-BSE, we obtained the eigen-energies, $\varepsilon_{e_1h_1e_2h_2}$ and $\varepsilon_{e_3h_3}$, and the wavefunctions, $B^{v_1,c_1,v_2,c_2}$ and $A^{v_3,c_3}$, for the 4B and 2B states, respectively. The dipole matrix elements of the possible 2B-4B transitions are then calculated as follows[62],

$$\langle e_3h_3|\hat{\mathbf{p}}|e_1h_1e_2h_2\rangle = \sum_{(v_i,c_i,i=1,2,3,4)} (A^{v_3,c_3})^* B^{v_1,c_1,v_2,c_2} \mathbf{p}^{v_4,c_4} \langle 0|\hat{a}^\dagger_{v_3}\hat{a}_{c_3}\hat{a}_{c_4}\hat{a}^\dagger_{v_4}\hat{a}^\dagger_{c_1}\hat{a}^\dagger_{c_2}\hat{a}_{v_1}\hat{a}_{v_2}|0\rangle$$

(3)

where $\hat{\mathbf{p}}$ is the dipole momentum operator defined via $\hat{\mathbf{p}} = \sum_{(v_4,c_4)} \mathbf{p}^{v_4,c_4} \hat{a}_{c_4}\hat{a}^\dagger_{v_4}$ and $\mathbf{p}^{v_4,c_4}$ is the independent-particle inter-band dipole momentum. The dielectric function $\epsilon$ can be calculated via

$$\epsilon \propto \frac{2\pi}{\hbar} \sum_\alpha \left| \sum_{(v_i,c_i,i=1,2,3)} \langle e_3h_3|\mathbf{e}\cdot\hat{\mathbf{p}}|e_1h_1e_2h_2\rangle_\alpha \right|^2 \Gamma(\varepsilon^\alpha_{e_1h_1e_2h_2} - \varepsilon_{e_3h_3} - \hbar\omega - i\gamma)$$

(4)

where, $\hbar\omega$ is the photon energy, **e** is the unit vector of the in-plane electric field component, **E**, for a normal incident field, $\alpha$ is the collective symbol for all the quantum numbers for the 4B eigen-states, and $\Gamma$ represents the Gaussian broadening function with a phenomenological broadening parameter, $\gamma$. The absorption coefficient is proportional to the imaginary part of the dielectric function. Such absorption coefficient corresponds to the excited state absorption, where the pump has generated 2B states such as various states of excitons and a subsequent weak probe generates a transition from 2B states to possible 4B states. Such a scenario corresponds to the typical pump-probe experiment[30,47] with an ultrafast delay.

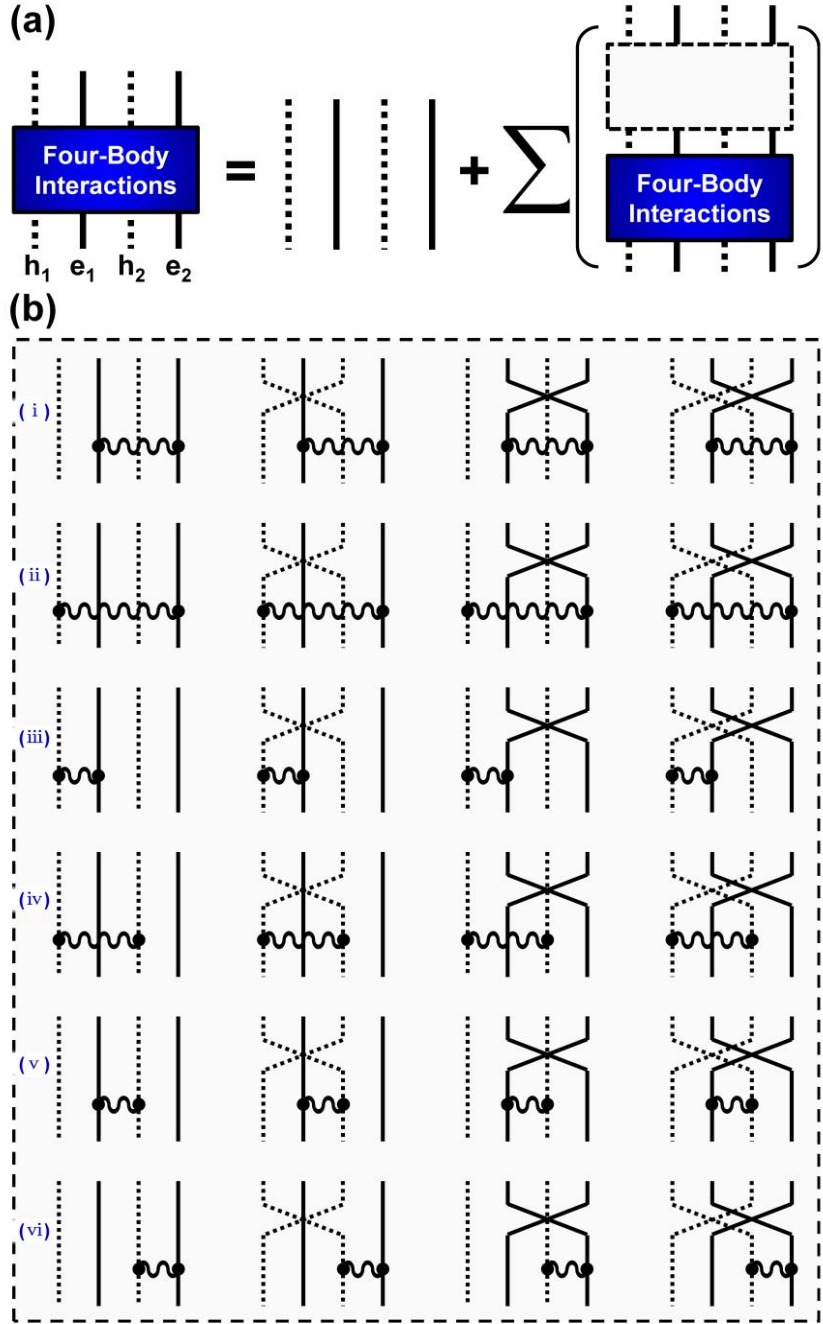

FIG. 1. The Feynman diagrammatic representations of the 2*e*2*h* 4B integral equation (a) and of the screened interaction kernels (labelled as (i) - (vi)) (b) which are part of the dashed box in (a). The dotted and solid lines represent the propagators of quasi-holes and electrons, respectively, where the crossing of the lines represent Fermionic permutation and the wavy lines with dotted ends represent the *e-h*, *e-e*, or *h-h* basic virtual scatterings. The unscreened interaction kernels are skipped here for brevity. For all the diagrams in (b), their signs (positive or negative) derived from the fermionic permutations are not specified.

Since in spectroscopic experiments T and X are easily-observed features, even with a simple CW experiment, they are typically used as references for identifying the new features[26-31]. We marked in FIG. 2 the spectral positions of T and X experimentally obtained from ML-MoTe$_2$[38,40,42,43]. The optical spectra calculated for ML-MoTe$_2$ using the 4B-BSE are shown in FIG. 2(a), (b) & (c), (d), corresponding to 2B-4B transitions with the 2B states are either the 1s-X or plasma. For the former case, there is a new peak ~10 meV below X. While for the case of plasma, there are likely more than six new peaks distributed from ~17 meV below T up to the region between T and X, covering a wide range of ~40 meV below X. We notice that the rich spectral features with so many identifiable peaks have not been seen before either in experiments or other theoretical studies related to BX[30,44-54] and BXFS[30,31], since similar four-particle equation has been studied only recently for BXFS[30,31] in W-based 2D materials. Another important observation from FIG. 2 is the difference between FIG.2(a), (c) and Figs. (b), (d), corresponding to two configurations co- and cross-polarizations between the pump and probe beams. The spectral features are much stronger and richer for the cross- than for the co-case, similar to the cases for bi-excitons[30,47].

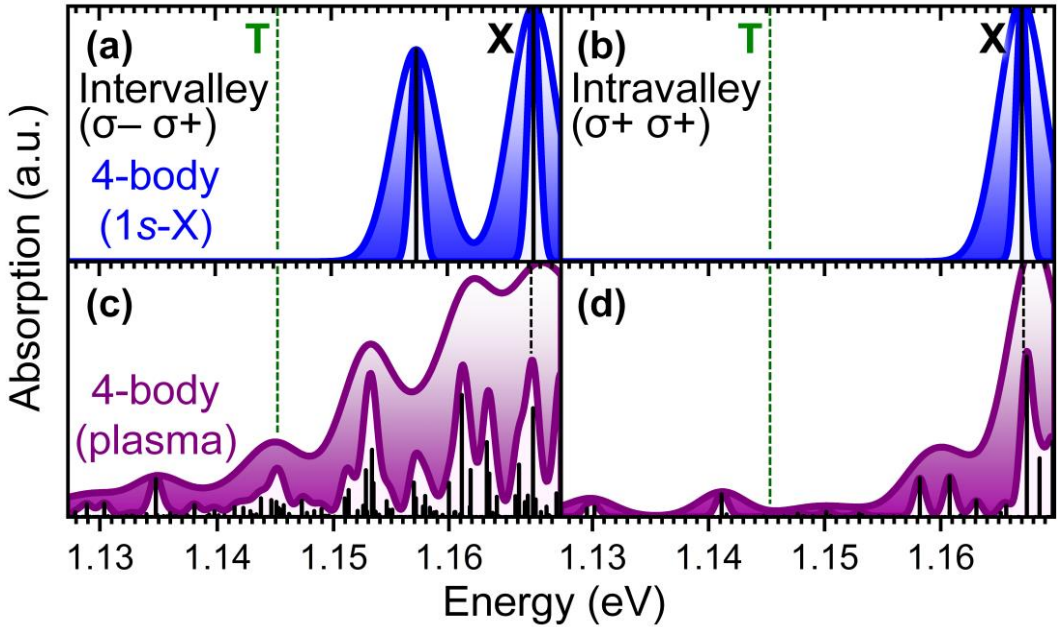

FIG. 2. Absorption spectra calculated based on the 4B-BSE for ML-MoTe$_2$. (a), (b), & (c), (d) show the spectra of transition between the 2B states given 1s-X ((a), (b)) and plasma ((c), (d)) and the 4B states solved from the full 4B-BSE. (a), (c), & (b), (d) were calculated in the way of assuming the 2B initial state created by pump at the K (σ–) or K' (σ+) valley and the probe field **E** to have a σ+ circular polarization. The vertical dashed lines mark the spectral energies of trion (T) and exciton typically

obtained from previous experiments on ML-MoTe$_2$. The vertical black lines underneath the spectral profiles mark the calculated spectral positions with the height representing the strength of the dipole transitions calculated from eq. (4). The spectral function, Γ, in eq. (4) is a Gaussian with a broadening parameter, γ, chosen to be 0.5 meV (unfilled profiles) or 2 meV (shaded profiles).

To shed more light into the origins of the new spectral features, we resort to the cluster expansion techniques in the following for its intuitive appeal. Following the diagrammatic representation of the 2B-BSE within the ladder approximation[58,59], we can also convert the self-consistent equation in FIG. 1(a) into a summation of infinite series as shown in FIG. 3(a). To achieve this, one could replace the blue box on the right in FIG. 1(a) with the entire right side of the equation, and repeat such replacement infinitely. In this way, the 4-particle correlation function could be transferred by such infinite iterations into the sum of a series of 4-propagator ladders, as shown in FIG. 3(a). Here, we show only the case for the intervalley 4B terms for brevity. The kernels in FIG. 1(b) should be the 1$^{st}$-order virtual scatterings, which could stack in various combinations and connect several of the 4 propagators to constitute higher-order virtual scatterings shown in FIG. 3(a). The total 4B interactions can be understood as the infinite summation of the various virtual scatterings (quadrupled ladders) of order 1, 2, …, ∞. These stacking ladders can be classified into several different partial summations. Consequently, each of the partial summations can be considered equivalent to the irreducible clusters of various orders (boxes in FIG. 3(a)) in the cluster expansion formalism (FIG. 3(b)).

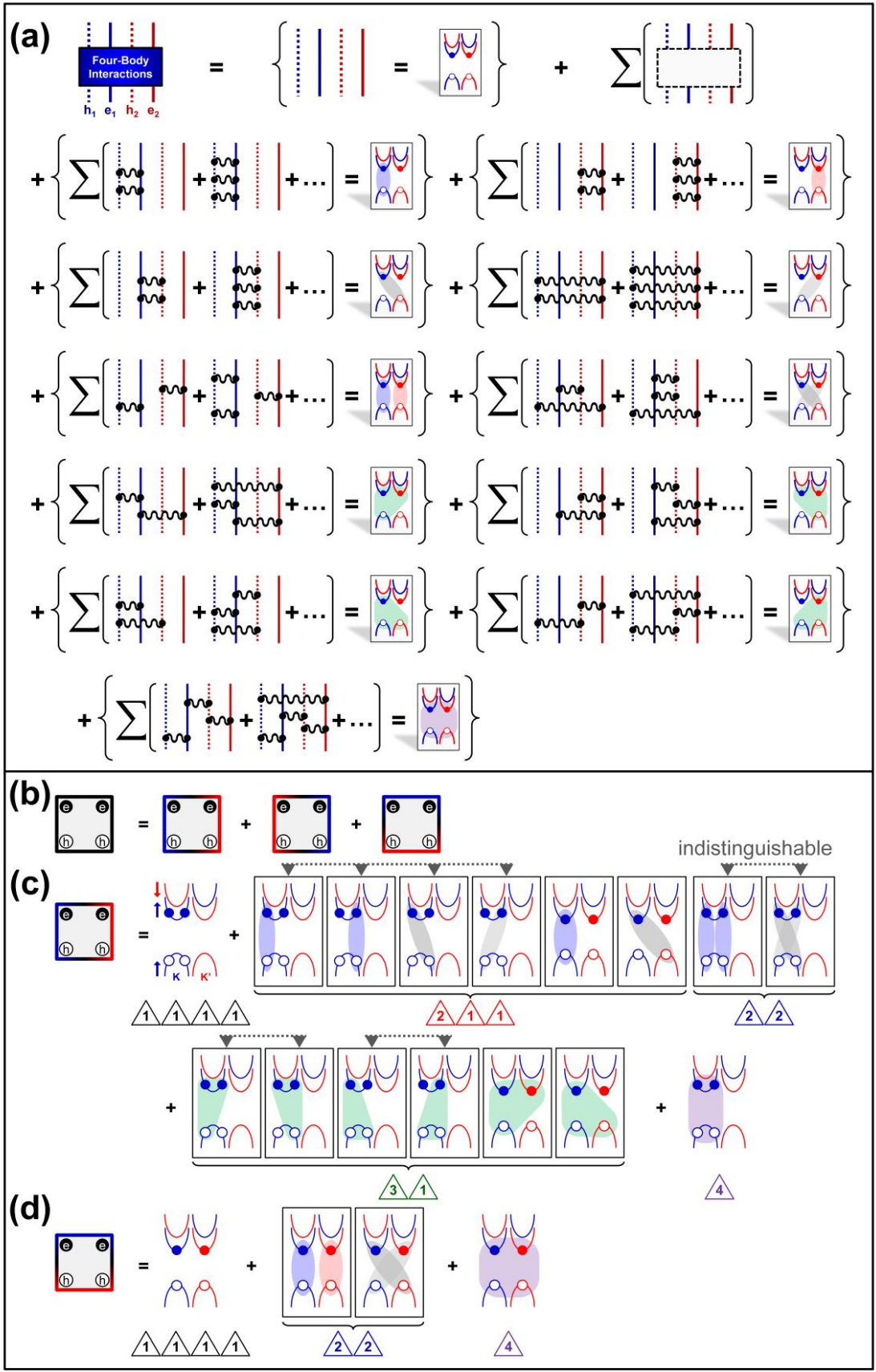

FIG. 3. Feynman diagrammatic representation of the summation of infinite series for the correlated 2e2h 4B system and the correspondences between the partial summations and the irreducible

clusters of various orders in the cluster expansion formalism (a). Even though all the inter- and intra-valley cases are included in the actual calculations, we only include the inter-valley cases in the figure for brevity, with the K- (spin-up) and K'- (spin-down) valley electrons colored in blue and red, respectively. Also for the sake of brevity of figures, the diagrams involving the unscreened interactions are not shown. (b) Decomposition of the 2$e$2$h$ 4B system into three parts according to the time reversal (K ↔ K') symmetry, as indicated by the blue-red symmetry of the square boxes. The first two terms/boxes are mutual partners under the time reversal and only one of them is presented in detail in (c). The third term/box is time reversal invariant, as presented in (d). (c) & (d) Cluster expansions of the 4B system into irreducible clusters (represented by the filled or unfilled circles connected by shaded shapes) of various orders (sizes) directly attached to the band structure for ML-MoTe$_2$ with valleys (K and K'). For brevity, we exclude those 2B clusters with the same charges, such as $e$-$e$ and $h$-$h$, in the figures, but these clusters are included in our theoretical calculations.

Specializing cluster expansion for individual electrons and holes including the spin and valley degrees of freedom, we can write down the complete sequence of irreducible clusters for the 2$e$2$h$ system as shown in FIG. 3(b) – (d). Here $\triangle_n$ represents an irreducible cluster of the n$^{th}$-order with n Fermions. Clearly $\triangle_1$ is a quasi-free electron or hole. $\triangle_2$ represents a direct or indirect $e$-$h$ pair (exciton), or a 2-body (2B) state, including all the excitonic Rydberg series[63]: 1$s$-X, 2$p$-X, 2$s$-X, …. $\triangle_3$ represents an $e$-$e$-$h$ or $e$-$h$-$h$ 3-body (3B) state. In the language of multiplons of Rausch $et~al$[1], $\triangle_2$, $\triangle_3$, and $\triangle_4$ are doublon, triplon, and quadruplon, …, respectively. Each time we include one more cluster of higher order into a truncated expansion, it re-introduces weak interactions (indicated by the wavy lines, ~) among irreducible clusters of the lower orders. Here, we truncated the cluster expansion up to the 4$^{th}$ order. As a result, the original non-interacting cluster $\triangle_2 \triangle_2$ shown in Fig. 3(c) & (d) representing two free excitons become weakly interacting, $i.e.$ $\triangle_2$ ~ $\triangle_2$, representing a bi-exciton or exciton-molecule, $e.g.$ (1$s$-X) ~ (1$s$-X) and all its excited states such as (1$s$-X) ~ (2$p$-X), (1$s$-X) ~ (2$s$-X) …. Obviously, cluster $\triangle_4$ or quadruplon (or quadron[64]) represents generally a distinct physical entity than the bi-exciton $\triangle_2$ ~ $\triangle_2$. The most fundamental difference between a bi-exciton and a quadruplon is the absence of an exciton in the latter and the lack of a clear association of any one of the two electrons to a given hole. But it is easy to imagine that a certain disassociation event (*e.g.* an excitation) of a quadruplon could possibly lead to the formation of a bi-exciton. In this sense, a bi-exciton can be considered an excited state of a quadruplon, which can thus be considered the ground state of all the 4B states.

Due to the intuitive pictures and transparent physics of the cluster expansion approach, we will use the language of cluster expansion to describe the results of our theory. For the 2B states, we chose 6 representative ones from a total of 15 discrete states of 1$s$-X, 2$p$-X, 2$s$-X, … plasma for ML-MoTe$_2$. For the 4B states, we consider following three cases of truncation, in order to extract and identify the effects of each: Case 1, up to $\triangle_2 \triangle_2$; Case 2, up to $\triangle_3 \triangle_1$; Case 3, up to $\triangle_4$. For each case, a series of 4B states are solved from the corresponding truncated 4B-BSE. To obtain a correct and complete picture of the possible 2B-4B transitions, we consider all the 2B and 4B states in the total energy scale as shown in FIG. 1 in SI S3 for various truncations.

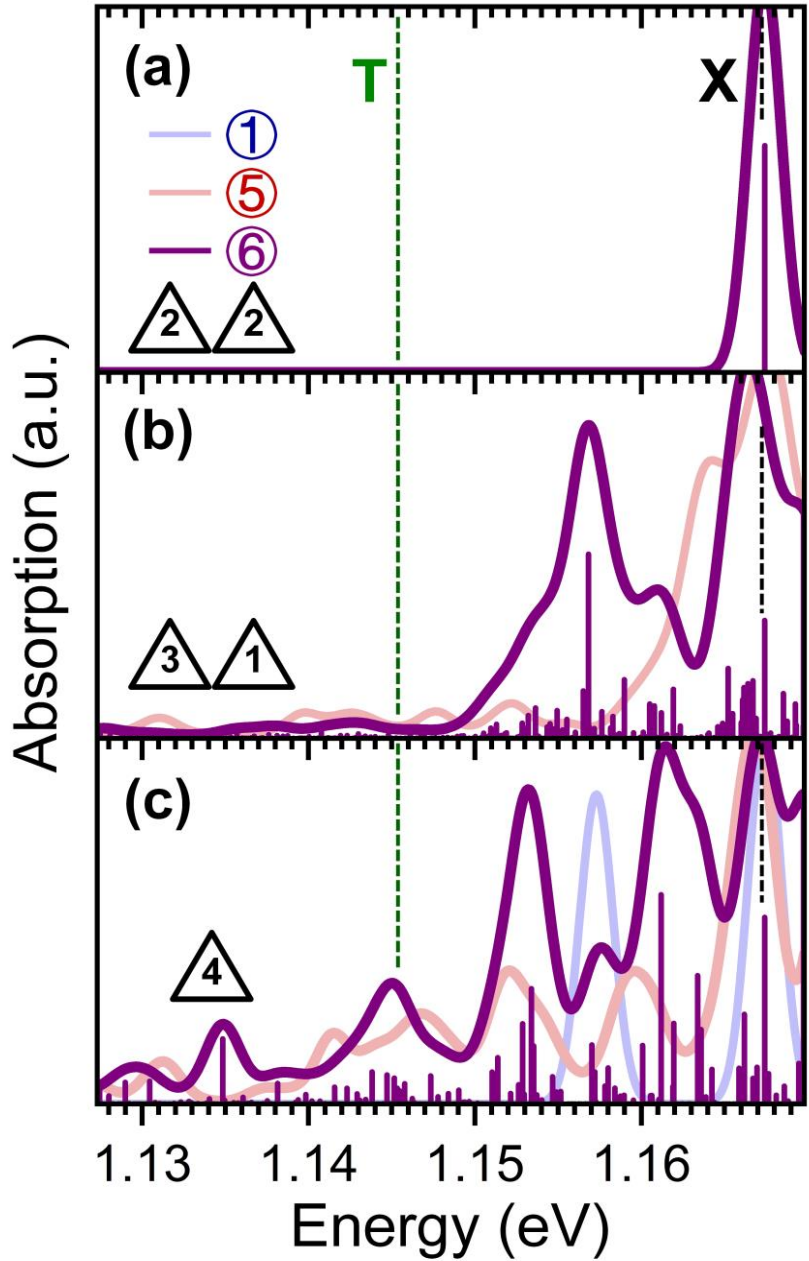

FIG. 4. 2B-4B transition spectra for various truncations. These figures are also enlarged versions of the lower left spectrum in FIG. 1(a), (b), and (c) in SI S3 with the corresponding alphabetic order. We choose from FIG. 1 in SI S3 only transitions ①, ⑤, ⑥ as representative ones to show here for brevity. The vertical dashed lines mark the spectral energies of T and X identical to those shown in FIG. 2. The broadening parameter, γ, is chosen to be 1.0 meV (profiles).

FIG. 4(a) & 4(b) show optical spectra obtained from the corresponding 2B-4B dipole transitions with the 4B-BSE truncated up to △₂△₂ and △₃△₁, respectively. The corresponding transitions in the total energy diagram are shown in Fig. 1(a) and 1(b) in SI S3. We notice that the main peak as well as some tiny splitting peaks between T

and X in the case of $\triangle_3\triangle_1$ (FIG. 4(b)) versus the featureless background in $\triangle_2\triangle_2$ (FIG. 4(a)). Strictly speaking, the only peak in FIG. 4(a) represents the 4B states of non-interacting clusters of type $\triangle_2\triangle_2$, such as (1s-X)(1s-X), (1s-X)(2s-X), ... (1s-X)(plasma). In the case of FIG. 4(b), $\triangle_3\triangle_1$ introduces interaction (wavy lines) between the low-order clusters, leading to the formation of $\triangle_2\sim\triangle_1\triangle_1$, $\triangle_2\sim\triangle_2$, ... Therefore, the main peak as well as those tiny splitting peaks below X in FIG. 4(b) should correspond to those of T⁻h & T⁺e ($\triangle_3\triangle_1$) or high-energy bi-excitons, e.g. likely (1s-X) ~ (plasma). Clearly, we do not see any spectral features emerging below T. This means that cluster $\triangle_2\triangle_2$, $\triangle_2\sim\triangle_2$, or $\triangle_3\triangle_1$ could not produce spectral peaks below T as shown in FIG. 2(c) or 4(c). Interestingly, cluster $\triangle_3\triangle_1$ should introduce coupling between $\triangle_2$ and $\triangle_2$ and thus converting them to $\triangle_2\sim\triangle_2$ or a bi-exciton. But such biexciton still does not introduce peaks below T.

Our next approximation is to include all the 4B states from the full 4B-BSE truncated up to $\triangle_4$. We see now features in FIG. 4(c) contains six or more spectral peaks, especially those below T, which clearly originate from cluster $\triangle_4$. By comparing various 2B-4B transitions in FIG. 4(c) for truncation up to $\triangle_4$, in FIG. 4(b) for $\triangle_3\triangle_1$, and in FIG. 4(a) for $\triangle_2\triangle_2$ both on the total-energy scale (see Fig. 1(c), 1(b), & 1(a) in SI S3) and optical spectrum, respectively, we could conclude the following: The states from cluster $\triangle_4$ for each of the same 2B states have lower energies and are thus more stable than those of $\triangle_2\sim\triangle_2$ (BX) and $\triangle_3\sim\triangle_1$ (T⁺~e and T⁻~h). Thus, our combined BSE and cluster expansion approach indicate the general existence of a more stable fourth order irreducible correlation, or quadruplon (quadron).

## Data availability
The data that support the findings of this study are available from the corresponding author upon request.

**Acknowledgements**

The authors acknowledge the following financial supports: National Natural Science Foundation of China (Grant No. 91750206, No. 61861136006); Pingshan Innovation Platform Project of Shenzhen Hi-tech Zone Development Special Plan in 2022(29853M-KCJ-2023-002-01); Universities Engineering Technology Center of Guangdong (2023GCZX005); Key Programs Development Project of Guangdong (2022ZDJS111); Tsinghua University Initiative Scientific Research Program.

Corresponding authors
Correspondence to Cun-Zheng Ning, ningcunzheng@sztu.edu.cn

**Ethics declarations**
Competing interests
The authors declare no competing interests.

# Supplementary Information

# Spectral Features of the Fourth Order Irreducible Correlations in a Monolayer Semiconductor

Jiacheng Tang,[1,2] Cun-Zheng Ning[2,1]*

[1]*Department of Electronic Engineering, Tsinghua University, Beijing 100084, China*
[2]*College of Integrated Circuits and Optoelectronic Chips, Shenzhen Technology University, Shenzhen 518118, China*

* Corresponding author. Email: ningcunzheng@sztu.edu.cn

**S1. Single-particle states**

To correct the DFT bands, we use the $G_0W_0$ quasi-particle bandgap of 1.77 eV[1] for ML-MoTe$_2$ by applying a ~0.82 eV move-up scissor operator to the original DFT gap of ~0.95 eV. In addition, we also used analytic expressions for the Coulomb interaction kernels[2-5]. Especially for the four-body calculations in this letter, the above approach could simplify the numerical computations greatly compared to an ab-initio DFT-GW-BSE method. Validity of such a reduction scheme has been verified and discussed in Ref. [3-5] for the 3B case.

**S2. Two-body and four-body Bethe-Salpeter Equations**

Coefficients $A^{v_3,c_3}$ and $B^{v_1,c_1,v_2,c_2}$ are determined by the 2B-BSE and 4B-BSE, respectively,

$$i\hbar\partial_t A^{v,c}_{k-Q,k}(0;Q) = \left(\tilde{\varepsilon}^{QP}_{ck} - \tilde{\varepsilon}^{QP}_{vk-Q} - i\gamma\right) A^{v,c}_{k-Q,k}(0;Q)$$

$$- \sum_{(v,c,q,k)} \left( W^{v',c,v,c'}_{k+q-Q,k,k-Q,k+q} - V^{v',c,c',v}_{k+q-Q,k,k+q,k-Q} \right) A^{v',c'}_{k+q-Q,k+q}(q;Q)$$

(1)

$$i\hbar\partial_t B^{v_1,c_1,v_2,c_2}_{k_1+q-Q,k_1,k_2-q,k_2}(0\,;Q)$$

$$= \left(\tilde{\varepsilon}^{QP}_{ck_1} + \tilde{\varepsilon}^{QP}_{ck_2} - \tilde{\varepsilon}^{QP}_{vk_1+q-Q} - \tilde{\varepsilon}^{QP}_{vk_2-q} - i2\gamma\right) B^{v_1,c_1,v_2,c_2}_{k_1+q-Q,k_1,k_2-q,k_2}(0\,;Q)$$

$$+ \sum_{(c'_1,c'_2,q')} \left\{ W^{c_1,c_2,c'_1,c'_2}_{k_1,k_2,k_1+q',k_2-q'} \left[ B^{v_1,c'_1,v_2,c'_2}_{k_1+q-Q,k_1+q',k_2-q,k_2-q'}(q'\,;Q) \right. \right.$$

$$\left. - B^{v_2,c'_1,v_1,c'_2}_{k_2-q,k_1+q',k_1+q-Q,k_2-q'}(q'\,;Q) \right]$$

$$- W^{c_1,c_2,c'_2,c'_1}_{k_1,k_2,k_2-q',k_1+q'} \left[ B^{v_1,c'_2,v_2,c'_1}_{k_1+q-Q,k_2-q',k_2-q,k_1+q'}(k_1-k_2+q'\,;Q) \right.$$

$$\left.\left. - B^{v_2,c'_2,v_1,c'_1}_{k_2-q,k_2-q',k_1+q-Q,k_1+q'}(k_1-k_2+q'\,;Q) \right] \right\}$$

$$- \sum_{(v'_1,c'_2,q')} \left( W^{v'_1,c_2,v_1,c'_2}_{k_1+q+q'-Q,k_2,k_1+q-Q,k_2+q'} \right.$$

$$\left. - V^{v'_1,c_2,c'_2,v_1}_{k_1+q+q'-Q,k_2,k_2+q',k_1+q-Q} \right) \left[ B^{v'_1,c_1,v_2,c'_2}_{k_1+q+q'-Q,k_1,k_2-q,k_2+q'}(q'\,;Q) \right.$$

$$- B^{v_2,c_1,v'_1,c'_2}_{k_2-q,k_1,k_1+q+q'-Q,k_2+q'}(q'\,;Q) - B^{v'_1,c'_2,v_2,c_1}_{k_1+q+q'-Q,k_2+q',k_2-q,k_1}(q'\,;Q)$$

$$\left. + B^{v_2,c'_2,v'_1,c_1}_{k_2-q,k_2+q',k_1+q+q'-Q,k_1}(q'\,;Q) \right]$$

$$- \sum_{(v'_1,c'_1,q')} \left( W^{v'_1,c_1,v_1,c'_1}_{k_1+q+q'-Q,k_1,k_1+q-Q,k_1+q'} \right.$$

$$\left. - V^{v'_1,c_1,c'_1,v_1}_{k_1+q+q'-Q,k_1,k_1+q',k_1+q-Q} \right) \left[ B^{v'_1,c'_1,v_2,c_2}_{k_1+q+q'-Q,k_1+q',k_2-q,k_2}(q'\,;Q) \right.$$

$$- B^{v_2,c'_1,v'_1,c_2}_{k_2-q,k_1+q',k_1+q+q'-Q,k_2}(q'\,;Q) - B^{v'_1,c_2,v_2,c'_1}_{k_1+q+q'-Q,k_2,k_2-q,k_1+q'}(q'\,;Q)$$

$$\left. + B^{v_2,c_2,v'_1,c'_1}_{k_2-q,k_2,k_1+q+q'-Q,k_1+q'}(q'\,;Q) \right]$$

$$+ \sum_{(v'_1,v'_2,q')} \left\{ W^{v'_1,v'_2,v_1,v_2}_{k_1+q+q'-Q,k_2-q-q',k_1+q-Q,k_2-q} \left[ B^{v'_1,c_1,v'_2,c_2}_{k_1+q+q'-Q,k_1,k_2-q-q',k_2}(q'\,;Q) \right. \right.$$

$$\left. - B^{v'_1,c_2,v'_2,c_1}_{k_1+q+q'-Q,k_2,k_2-q-q',k_1}(q'\,;Q) \right]$$

$$- W^{v'_2,v'_1,v_1,v_2}_{k_2-q-q',k_1+q+q'-Q,k_1+q-Q,k_2-q} \left[ B^{v'_2,c_1,v'_1,c_2}_{k_2-q-q',k_1,k_1+q+q'-Q,k_2}(k_1-k_2+2q-q'\right.$$

$$- Q\,;Q)$$

$$\left.\left. - B^{v'_2,c_2,v'_1,c_1}_{k_2-q-q',k_2,k_1+q+q'-Q,k_1}(k_1-k_2+2q-q'-Q\,;Q) \right] \right\}$$

$$-\sum_{(v'_2,c'_1,q')} \left(W^{v'_2,c_1,v_2,c'_1}_{k_2-q+q',k_1,k_2-q,k_1+q'}\right.$$

$$\left. - V^{v'_2,c_1,c'_1,v_2}_{k_2-q+q',k_1,k_1+q',k_2-q}\right) \left[B^{v_1,c'_1,v'_2,c_2}_{k_1+q-Q,k_1+q',k_2-q+q',k_2}(q';Q)\right.$$

$$- B^{v'_2,c'_1,v_1,c_2}_{k_2-q+q',k_1+q',k_1+q-Q,k_2}(q';Q) - B^{v_1,c_2,v'_2,c'_1}_{k_1+q-Q,k_2,k_2-q+q',k_1+q'}(q';Q)$$

$$\left. + B^{v'_2,c_2,v_1,c'_1}_{k_2-q+q',k_2,k_1+q-Q,k_1+q'}(q';Q)\right]$$

$$-\sum_{(v'_2,c'_2,q')} \left(W^{v'_2,c_2,v_2,c'_2}_{k_2-q+q',k_2,k_2-q,k_2+q'}\right.$$

$$\left. - V^{v'_2,v_2,c'_2,c_2}_{k_2-q+q',k_2-q,k_2+q',k_2}\right) \left[B^{v_1,c_1,v'_2,c'_2}_{k_1+q-Q,k_1,k_2-q+q',k_2+q'}(q';Q)\right.$$

$$- B^{v'_2,c_1,v_1,c'_2}_{k_2-q+q',k_1,k_1+q-Q,k_2+q'}(q';Q) - B^{v_1,c'_2,v'_2,c_1}_{k_1+q-Q,k_2+q',k_2-q+q',k_1}(q';Q) +$$

$$\left. B^{v'_2,c'_2,v_1,c_1}_{k_2-q+q',k_2+q',k_1+q-Q,k_1}(q';Q)\right]$$

(2)

In eq. (1) & (2) above, $\tilde{\varepsilon}^{QP}_{ck}$ and $\tilde{\varepsilon}^{QP}_{vk-Q}$ stand for the corrected band energies. Additionally, a homogeneous broadening factor, $\gamma$, is used to mimic the finite excited-state lifetime. $W^{v',c,v,c'}_{\bar{k}',k,\bar{k},k'}$ and $V^{v',c,c',v}_{\bar{k}',k,k',\bar{k}}$ represent the screened and unscreened (bare) Coulomb interaction kernels and are defined as $W^{v',c,v,c'}_{\bar{k}',k,\bar{k},k'} = \int \psi^{v'}_{\bar{k}'}(r)^* \psi^{c}_{k}(r')^* w(r,r') \psi^{v}_{\bar{k}}(r) \psi^{c'}_{k'}(r') d^d r d^d r'$ and $V^{v',c,c',v}_{\bar{k}',k,k',\bar{k}} = \int \psi^{v'}_{\bar{k}'}(r)^* \psi^{c}_{k}(r')^* v(r,r') \psi^{c'}_{k'}(r) \psi^{v}_{\bar{k}}(r') d^d r d^d r'$ (d represents the dimension), respectively, via the screened Coulomb potential $w(r,r')$ and the unscreened one $v(r,r')$[2-4,6-12].

To simplify the computation, we used the zero differential overlap approximation[2] to generate those Coulomb interaction kernels in the reciprocal space as functions of the momentum transfer, q. We used the Rytova-Keldysh[13,14] formula $W_q = -\frac{2\pi e^2}{q(1+2\pi\chi_{2D}q)}$ as the screened Coulomb potentials ($W_q$) and $V_q = -\frac{2\pi e^2}{q}$ as the unscreened one ($V_q$)[2-5]. Validity and accuracy of such an approximation scheme has been verified in Ref.

[2-5] for those 2B and 3B calculations. It has been shown[3,4,8,10-12] that any *e-e*/*h-h* term (including the corresponding Fermi permutations) must be a screened Coulomb interaction kernel, while any *e-h* term should include both the screened and unscreened interactions. It is exactly the latter, the unscreened *e-h* interaction kernels that account for the *e-h* exchange effect and are involved in producing the fine structures of trions[8,10,15-17] and bi-excitons[11,12]. Since the absences of those unscreened kernels would not distort the understanding of the 4B interaction picture, we skipped them and depicted only the screened ones in FIG. (1) & (3) in the main text. However, we point out that we included all of the interaction kernels completely in our theoretical calculation.

**S3. 2B-4B transitions shown in the total-energy scale**

To obtain a correct and complete picture of the possible 2B-4B transitions, we consider all the 2B and 4B states in the total energy scale as show in FIG. 1 in this section for various truncations. A series of transitions occur between different 2B states and the corresponding 4B states on the total energy scale, as marked by the vertical double-arrowed lines with ① – ⑥ (6 representative examples). It is important to emphasize that the optical spectrum measured in an absorption or emission experiment contains only the energies differences between the two states. Therefore, all these transitions are shifted relative to a common reference (*e.g.* the 1*s*-X energy), as shown along with the direction of horizontal axis. The actual optical spectrum is the superposition (left lower corner) of all of these spectra shown along with the direction of horizontal axis, or all vertically "collapsed" spectra shown along with the direction of vertical energy axis.

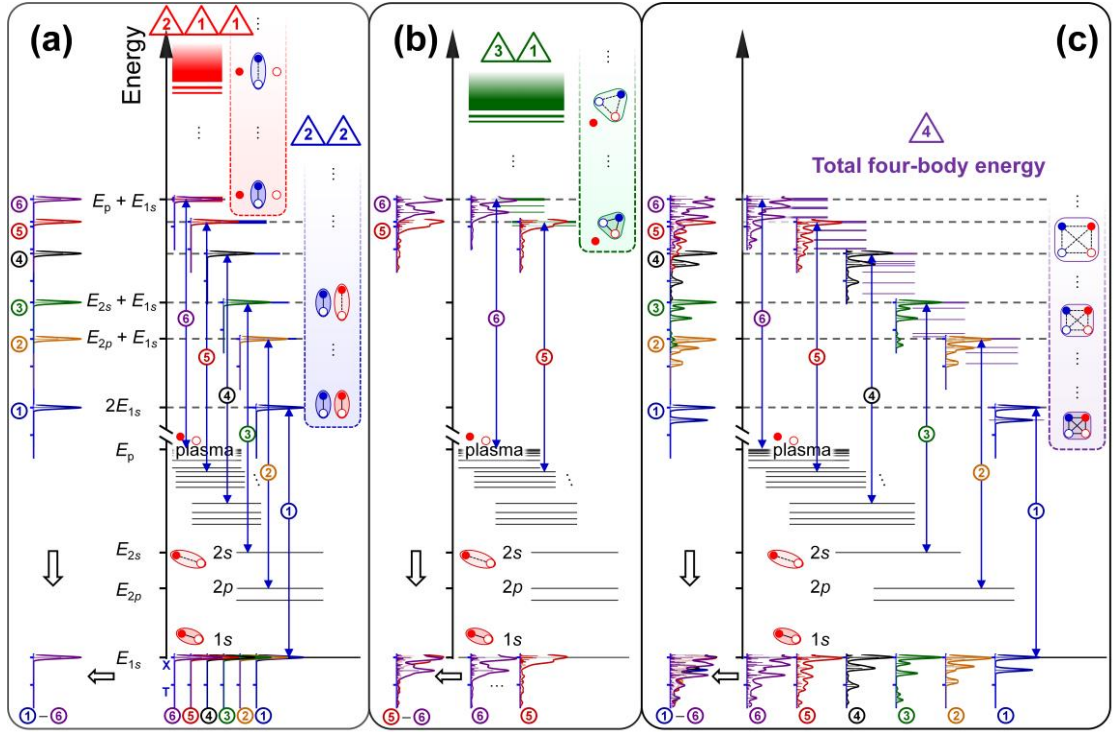

FIG. 1. Schematic of the correspondences between the 2B-4B optical transition spectra and the total energy spectra. (a), (b), (c) correspond to the cases of the 4B system being truncated up to the irreducible clusters of different orders, i.e. △△ (a), △△ (b) and △ (c). Since the calculations for T⁻h (shown in b) and its counterpart, T⁺e, gave the identical results, we illustrated only the former in (b) for brevity. See the text in this section for detailed descriptions of the figures.

Importantly, we notice from the transitions in the total-energy scale that the lower-frequency features (below T) in the overlapped optical spectrum correspond to the transitions (such as ④, ⑤, and ⑥) between the more-excited states in the 4B manifold. Moreover, only the transitions between the highly excited 2B and 4B states such as those of ③ – ⑥, can contribute to the peaks well below T. This could suggest that if a 4B quantum kinetic is further calculated or measured, the features below T could decay faster than those between T and X. In addition, nearly all the 2B-4B transitions shown in the total-energy spectrum, *e.g.* from ① (low energy) to ⑥ (high energy), can contribute to the spectral peaks between T and X. Especially, the spectral peak related to the 4B ground state is calculated to be between T and X. This could suggest that in actual overlapped spectra, intensities of the features between T and X should be stronger than those below the T peak. For all the spectra calculations, we determined the single-particle basis by truncating the momentum space as a circle around the K and K' points to further reduce the computation. We found the results, e.g. spectral energies or relative energy differences, could converge well within ~2 meV using a k-mesh of ~36×36×1 and a truncation radius of 0.227 Å$^{-1}$. Similar parameters were used for calculating bi-exciton binding-energies[11,12].